\documentclass[12pt]{article}
\usepackage{amsfonts}
\usepackage{amsmath}
\usepackage[colorlinks=true,citecolor=blue]{hyperref}
\begin{document}
	\title{Fission with a difference}
	\author{G. S. Burra\\
		Adj. Professor, University of Udine, 201, Via delle Scienze\\
		33100, Udine, Italy}
	\date{}
	\maketitle
	\begin{abstract} Invoking a model of an elementary particle as a collection 
	of ultrarelativistic transient particles we show that it is possible to 
	recover the energy of the particle by bombarding it with monochromatic 
	high-energy radiation.
	\end{abstract}

\section*{Introduction} We consider the possibility of using an alternative 
route to releasing fission energy. This is prompted by some recent developments 
by the team of scientists Cruz Chu et. al. \cite{cruz-21}, which leads to the 
technological development of practically monochromatic radiation in the X-ray 
region.\\
Let us start from   a relativistic point of view, and the Lorentz
transformation,
\begin{equation}
	x = \gamma (x'- v t), \gamma = (1 - v^2/c^2)^{-1/2}\label{e28}
\end{equation}

Indeed it is known that for a collection of relativistic particles, the
various mass centres form a two-dimensional disc perpendicular to the
angular momentum vector $\vec L$ and with radius (ref.\cite{r32})
\begin{equation}
	r = \frac{L}{mc}\label{e29}
\end{equation}
Further if the system has positive energies, then it must have an
extension greater than $r$, while at distances of the order of
$r$ we begin to encounter negative energies.\\
If we consider the system to be a particle of spin or angular momentum
$L=\frac{\hbar}{2}$, then equation (\ref{e29}) gives, $r = \frac{\hbar}{2mc}$.
That is we are in the Compton wavelength region. Another interesting feature
which is the two dimensionality of the  disc
of mass centres.\\
On the other hand it is known that (cf. ref.\cite{r20}), if a Dirac particle
is represented by a Gausssian packet, then we begin to encounter negative
energies precisely at the same  Compton wavelength as above. Thus a
particle can indeed be treated as  a spherical shell of
relativistic transient sub constituents or ``particlets." Indeed, this is an 
alternative description of Dirac's zitterbewegung or rapid oscillation. \\
The above picture is also reminiscent of Dirac's shell or membrane model of
the electron\cite{r33,r34,r35}.\\
 Outside this Compton region we
have the usual space (or space time) of physics. But as we approach the
Compton wavelength region we encounter a region where
the space axis becomes as it were a complex plane. This has been described at 
length by the author, in terms of the Feschbach formalism \cite{fesch} which 
leads to the double Weiner process.
 Consider the following system 
 \begin{equation} \label{eq:fesch} 
 	\begin{split}
 	i \hbar\frac{\partial \phi}{\partial t} & = \frac{1}{2m} 
 	\left(\frac{\hbar}{i 
 	\bf{\nabla}}-\frac{e \bf{A}}{c}\right)^2(\phi+\chi)+(e\phi+m c^2)\phi \\
 	i \hbar\frac{\partial \phi}{\partial t} & = -\frac{1}{2m} 
 \left(\frac{\hbar}{i 
 	\bf{\nabla}}-\frac{e \bf{A}}{c}\right)^2(\phi+\chi)+(e\phi-m c^2)\phi
 	\end{split}
 \end{equation}
\cite{bgs-tdu-08}.) The merit of this formalism is that it enables us to give a 
particle interpretation to the usual wave-formalism  (see \cite{fesch} for 
further details.) However the advantage of the Feschbach Villars formalism is 
that we can now work with an ostensible particle interpretation. \\
In any case we encounter the Compton scale again and again. Wigner \cite{r36}
pointed out its remarkable universality.\\
From the above it is apparent that if an elementary particle in the above 
characterisation is bombarded with very high frequency  radiation of the order 
of the Compton frequency such a particle would break up and yield its energy.
What happens in this case is  that the 
bell curve becomes so compressed that it will be like a straight line or spike, 
almost (see  \cite{berry-08,kos-08}.) This sharp spike would 
break up the elementary particle releasing it’s Mass as energy.
\\ It is well known in Quantum Mechanics that,
what may be called monochromatic waves are
an idealization. This is in the sense that we have in general 
a wave packet made up of several frequencies
\cite{PoCr-61}. But
suppose we can single out a pure or nearly pure
frequency ? This is a technological problem.
Let us start with the Schrodinger equation:\cite{PoCr-61}
$$ \frac{d^2 \psi}{d x^2}+\frac{p^2}{\hbar^2} \psi = 0 $$
where $$ p= \sqrt{ 2m[E-V(x)]}.$$
This leads to \begin{equation} \label{eq:3}  \phi(x) e^{\pm 
\frac{i}{\hbar}\int^x p(x) dx  } 
\end{equation}
and we already have a wave packet over different
values of $p$ or effectively frequencies. However
if we have a wave function like $\psi'=e^{i k x -pt}$,
such a wave would be an extreme idealization
and at the same time would  monochromatic. Can we
achieve this, is the question. There has been 
recently some progress in this direction thanks
to the experiment of Cruz-Chu and Co-workers \cite{cruz-21}
who have been able to conduct an experiment
where single particle X ray diffraction patterns
could be analysed thanks to a machine learning
algorithm.
\section*{Remarks} What happens in this case is, the bell curve becomes so 
compressed that it will be like a 
straight line. This 
sharp spike could break up the elementary particle releasing it’s mass as 
energy. Fortunately, in recent years there has been some progress in this 
direction \cite{rod-21,berry-08,kos-08}. Furthermore, it maybe pointed out that 
a pure monochromatic signal would be useful in communications as well. This is 
because, effectively the bandwidth would increase \cite{bgs-NAP-21}.
Finally, we observe that, if we can break up quarkonium particles, we can 
extract even greaterr energy. There is one way of doing this: we know that with 
$g=2$ factor, there is a sort of precession and, if we could radiate with 
resonant frequencies, the particle would break up. This could be a 
technological problem.
\section*{Appendix} We use some recent significant experimental results. The 
first was, the 
containment of plasma, roughly protons and other particles at a very high 
temperature came about in joint US – UK experiments.
The next has been laser 
cooling techniques used to prepare short range forces. Further the experimental 
isolation of monochromatic waves, and finally  the author's formulation of 
quantum mechanics in terms of beats and also his work on Bose-Einstein 
condensation at a temperature $> p=0.$  All these together give a sense that  
fusion can be done at lower temperatures. This could be a breakthrough in 
itself rather like Bose-Einstein condensation at is slightly higher temperature.

\end{document}